\documentclass[prd,nofootinbib]{revtex4}

\usepackage[mathscr]{euscript}
\usepackage{dcolumn}
\usepackage{bm}
\usepackage{graphicx}
\usepackage{subfigure}
\usepackage{dcolumn}
\usepackage{multirow}
\usepackage{amsmath,amssymb,amsthm}
\usepackage{mathtools,leftidx}
\usepackage{tensor}
\usepackage[colorlinks=true,linkcolor=blue]{hyperref}

\newcommand{\rmnum}[1]{\uppercase\expandafter{\romannumeral #1\relax}}
\newcommand{\bea}{\begin{eqnarray}}
\newcommand{\eea}{\end{eqnarray}}
\newcommand{\beq}{\begin{equation}}
\newcommand{\eeq}{\end{equation}}
\newcommand{\nn}{\nonumber}
\def\/{\over}

\begin{document}
	
\title{Joint constraint on primordial gravitational waves and polarization rotation angle with current CMB polarization data }

\author{Hua Zhai$^{1,2}$, Si-Yu Li$^{3}$, Mingzhe Li$^{4}$, and Xinmin Zhang$^{1, 2}$}

\affiliation{$^1$Theoretical Physics Division, Institute of High Energy Physics (IHEP), Chinese Academy of Sciences, 19B Yuquan Road, Shijingshan District, Beijing 100049, China}
\affiliation{$^2$University of Chinese Academy of Sciences, Beijing, China}
\affiliation{$^3$Key Laboratory of Particle Astrophysics,  Institute of High Energy Physics (IHEP), Chinese Academy of Sciences, 19B Yuquan Road, Shijingshan District, Beijing 100049, China}
\affiliation{$^4$Interdisciplinary Center for Theoretical Study, University of Science and Technology of China, Hefei, Anhui 230026, China}

\begin{abstract}
	
Cosmological CPT violation will rotate the polarized direction of CMB photons, convert partial CMB E mode into B mode and vice versa. 
It will generate non-zero EB, TB spectra and change the EE, BB, TE spectra.
This phenomenon gives us a way to detect the CPT-violation signature from CMB observations, and also provides a new mechanism to produce B mode polarization. In this paper, we perform a global analysis on tensor-to-scalar ratio $r$ and polarization rotation angles based on  current CMB datasets with both low $\ell$  (Planck, BICEP2/Keck Array) and high $\ell$ (POLARBEAR, SPTpol, ACTPol).
Benefited from the high precision of CMB data, we obtain the isotropic rotation angle $\bar{\alpha} = -0.01^\circ \pm 0.37^\circ $ at 68\% C.L.,  the variance of the anisotropic rotation angles $C^{\alpha}(0)<0.0032\,\mathrm{rad}^2$, the scale invariant power spectrum $D^{\alpha\alpha}_{\ell \in [2, 350]}<4.71\times 10^{-5} \,\mathrm{rad}^2$ and  $r<0.057$ at 95\% C.L.. Our result shows that with the polarization rotation effect, the 95\% upper limit on $r$ gets tightened by 17\%. 

\end{abstract}

\maketitle

\section{introduction}

In the standard model of particle physics, the Charge-Parity-Time Reversal (CPT) symmetry is an exact symmetry and occupies a fundamental status. So probing CPT violation is an important way to test the standard model, and a very useful approach to searching for the new physics. Up to now, CPT symmetry has passed a number of high-precision experimental tests and no definite signal of its violation has been observed in the laboratory. So, the  CPT violation, if exist, should be very small to be amenable to the laboratory experimental limits.

However, the CPT symmetry could be dynamically broken in the expanding universe.  For instances, in Refs. \cite{Cohen:1987vi, Li:2001st,Li:2002wd,Davoudiasl:2004gf,Li:2004hh}, the cosmological CPT violation has been considered to generate the baryon number asymmetry in the early universe. A notable property of this kind of baryogenesis models is that the CPT violation at present time is too small to be detected by the laboratory experiments, but large enough in the early universe to account for the  observed baryon number asymmetry.  As shown in Refs. \cite{Feng:2004mq,Feng:2006dp,Li:2006ss}, such type of CPT violations might be observed by the cosmological probes. With the accumulation of high-quality observational data, especially those from the cosmic microwave background (CMB) experiments, cosmological observation becomes a powerful way to test CPT symmetry.  

Generally the cosmological CPT violation in the photon sector can be modeled by the coupling between photons and an external field $\theta(x)$ through the Chern-Simons Lagrangian,
\bea\label{cs}
\mathcal{L}_{cs} = \theta(x) F_{\mu\nu} \widetilde{F}^{\mu\nu}~,
\eea
where $F_{\mu\nu}=\partial_{\mu}A_{\nu}-\partial_{\nu}A_{\mu}$ is the electromagnetic tensor and  $\widetilde{F}^{\mu\nu}=(1/2)\epsilon^{\mu\nu\rho\sigma}F_{\rho\sigma}$ is its dual. If  $\theta$ is a constant, the Chern-Simons term will have no effect on the dynamics of photons because the Pontryagin density $F_{\mu\nu} \widetilde{F}^{\mu\nu}$ is a total derivative and the vacuum here is topological trivial. There are at least two approaches to get $\theta(x)$ as a variable. With the first approach, $\theta(x)=p_{\mu}x^{\mu}$ is constructed by a non-dynamical vector $p_{\mu}$. However when considering the couplings to gravity, this case is not compatible with general relativity and its covariant extensions \cite{Li:2009rt}. In the second approach, $\theta(x)=f(\phi(x))$,where $f(\phi(x))$ is a general function of a dynamical scalar field $\phi(x)$.  Such a scalar field may be the dynamical dark energy \cite{Wetterich:1987fm,Caldwell:1999ew,Feng:2004ad} as in Refs. \cite{Li:2001st,Li:2002wd}, or axion-like field, or the curvature of spacetime \cite{Davoudiasl:2004gf,Li:2004hh}. During the evolution of the universe,  $\theta(x)$ is treated as an external field, its evolution or configuration picks up a preferred frame, so that the Chern-Simons term (\ref{cs}) is not invariant under Lorentz and CPT transformations. The physical consequence of this CPT violation is to cause the rotations of the polarization directions of photons when propagating in the space. This holds for both the homogeneous and  inhomogeneous universe \cite{Li:2008tma}. In terms of the Stokes parameters for linear polarized photons, the rotation can be expressed as 
\beq\label{qu}
 \widetilde Q \pm i \widetilde  U = \exp(\pm	i 2\alpha)(Q\pm i U)~,
\eeq
where the rotation angle is twice the integral of $\partial_{\mu}\theta$ along the light ray from the source to the observer,
\bea
\alpha =2 \int_s^o  \partial_{\mu}\theta dx^{\mu}=2[\theta(x_o)-\theta(x_s)]~,
\eea
and finally depends on the difference of $\theta$ between the observer and the source. 
For CMB, the source is on the last scattering surface, when setting the observer at the origin point of the coordinate system, the rotation angle is generally expressed as 
\bea\label{alpha}
\alpha(\bm{n})=-2\theta(\Delta\eta \cdot \bm{n})~,
\eea
where $\bm{n}$ represents the direction of the point at the last scattering surface and $\Delta \eta=\eta_0-\eta_{lss}$ is the conformal time difference of the present time to the last scattering surface. 
The rotation has the ability to convert part of E-mode polarization to B-mode polarization, and vice versa. This will change the power spectra of CMB polarization, especially induce nonzero TB and EB spectra \cite{Lue:1998mq,Feng:2006dp}. Such effects give a  way to detect or constrain the rotation angle, then the CPT-violation signature with CMB data.

The rotation angle $\alpha(\bm{n})$ is generally a direction dependent scalar field on 2-d sphere, as shown in Eq. (\ref{alpha}). It is originated from the dynamics of the scalar field $\phi(x)$ and is expected to have position dependent fluctuations due to the dynamical feature.  It is natural to split  $\alpha(\bm{n})$ into the isotropic and the anisotropic parts $
\alpha(\bm{n})=\bar{\alpha} + \delta\alpha(\bm{n})$ as we have usually done in the cosmological perturbation theory.  The isotropic rotation angle, $\bar{\alpha}$, can be considered as the mean of $\alpha(\bm{n})$ over the sphere.  At the leading order, we may only consider the isotropic rotation angle as an approximation.  For this case, the rotated CMB power spectra have simple forms \cite{Feng:2006dp}, the details will be presented in the next section. With WMAP and BOOMERANG (B03) data, Feng et.al \cite{Feng:2006dp} has performed the first measurement on rotation angle. Since then, a lot of works have been done in terms of the observed CMB polarization data along this line.
The isotropic rotation angle $\bar{\alpha}$ has been constrained by various collaborations of CMB surveys, including QUaD\cite{Wu:2008qb}, WMAP\cite {Hinshaw:2012aka}, ACTPol\cite{Louis:2016ahn} and Planck\cite{Aghanim:2016fhp}, and by combined datasets including CMB and LSS observations \cite{Xia:2009ah, Zhao:2015mqa, Xia:2012ck}. The constraint on $\bar{\alpha}$ in these works is found to be at level of one degree.
 
A comprehensive study on the Chern-Simon effect should include the anisotropies of the rotation angle\cite{Li:2008tma}. If the anisotropies are random and satisfy the Gaussian statistics, they can be described fully by an angular power spectrum $C_{\ell}^{\alpha\alpha}$ .  In terms of $C_{\ell}^{\alpha\alpha}$,  Ref. \cite{Li:2008tma, Li:2013vga} derived the formulae of distortion effect on CMB power spectra in a non-perturbative way, and then Ref. \cite{Li:2013vga, Li:2014oia} constrained the anisotropies by global fitting to the combination of CMB observations.
Anisotropic rotation angles are also studied in  Ref.\cite{Kamionkowski:2008fp, Yadav:2009eb, Gluscevic:2009mm}  using the four point correlation function  method.  In this way, the constraints on the anisotropies of rotation angle with WMAP-7\cite{Gluscevic:2012me}, POLARBEAR\cite{Ade:2015cao}, and BICEP2/Keck Array\cite{Array:2017rlf} were obtained.
Up to now, these results showed that the anisotropies of the rotation angle, if exist, should be very small.   

On the other hand, rotation angle also provides another mechanisms of producing CMB B-mode polarization, so that it will affect the detection of PGW. 
Authors in Ref. \cite{Li:2014cka} studied the degeneracy between $r$ and the isotropic rotation angle.
Authors in Ref. \cite{Li:2014oia, Li:2015vea} studied the degeneracy between $r$ and the anisotropic rotation angles through a global fitting to combination data of BK14 and Planck 2015. They found the peak feature in the one-dimensional probability distribution of $r$ reported in BICEP2/Keck Array collaboration paper\cite{Ade:2018gkx} disappeared after taking rotation angle effect into account.  

In this paper, we perform a comprehensive global fitting on $r$ and the isotropic/anisotropic rotation angles with current CMB data. 
Besides the low $\ell$ CMB data (PLANK\cite{Aghanim:2018eyx}, BICEP2/Keck Array\cite{Ade:2018gkx}), we also adopt the latest CMB data with high $\ell$ such as the POLARBEAR\cite{Ade:2014afa}, SPTpol\cite{Henning:2017nuy, Keisler:2015hfa}, ACTPol\cite{Louis:2016ahn}. Our results show  that isotropic and anisotropic rotation angles generate B mode in different ways. In general, the BB spectrum, especially at small scales, is more sensitive to the anisotropic angles than the isotropic one.

This paper is organized as follows: in Section \ref{section2}  we briefly review the effects of polarization rotation on the CMB power spectra and introduce the basic formulae for numerical calculations; in Section \ref{section3} we describe the datasets adopted in this work and the numerical result of the global analysis; Section \ref{summary} dedicates to the summary.
 
\section{ROTATED CMB polarization POWER SPECTRA}
\label{section2}

In this section, we first briefly review the effects of polarization rotations induced by CPT violation on CMB. Then we will introduce the basic formulae used for numerical calculations in this paper. 

The rotated CMB power spectra can be obtained by applying Eq. (\ref{qu}) to the angular power spectra. For the isotropic rotation $\bar{\alpha}$, the rotated spectra take the following form \cite{Feng:2006dp}:  

\bea\label{iso_rotation}
\widetilde{C}^{TB}_{\ell}&=&C^{TE}_{\ell} \sin (2\bar{\alpha})~,\nonumber\\
\widetilde{C}^{EB}_{\ell}&=&\frac{1}{2}(C^{EE}_{\ell}-C^{BB}_{\ell}) \sin (4\bar{\alpha})~,\nonumber\\
\widetilde{C}^{TE}_{\ell}&=&C^{TE}_{\ell} \cos (2\bar{\alpha})~,\nonumber\\
\widetilde{C}^{EE}_{\ell}&=&C^{EE}_{\ell} \cos^2 (2\bar{\alpha})+C^{BB}_{\ell} \sin^2 (2\bar{\alpha})~,\nonumber\\
\widetilde{C}^{BB}_{\ell}&=&C^{BB}_{\ell} \cos^2 (2\bar{\alpha})+C^{EE}_{\ell} \sin^2 (2\bar{\alpha})~,
\eea

More generally, we should consider fluctuation of the rotation angle $\delta\alpha(\bm{n})$, which gives rise to further distortions to the power spectra.  To start with, we expand $\delta\alpha(\bm{n})$ in series of spherical harmonics:
\begin{eqnarray}
\delta\alpha(\bm{n})&=&\sum_{\ell,m}{\alpha_{\ell m}Y_{\ell m}}(\bm{n})~,
\end{eqnarray}
where $\alpha_{\ell m}$ denotes the corresponding coefficient of the spherical harmonic expansion. 
If $\delta\alpha(\bm{n})$ is a Gaussian random field, it is fully described by the two-point correlation function $C^{\alpha}(\beta)$ in the pixel space or the power spectrum $C^{\alpha\alpha}_{\ell}$ in the angular space. They are related through the following equation,
\begin{equation}
C^{\alpha}(\beta)\equiv\left(\delta\alpha(\bm{n})\delta\alpha(\bm{n}')\right) = \sum_{\ell} {\frac{2\ell+1}{4\pi}C_{\ell}^{\alpha\alpha}P_{\ell}(\cos\beta)}~,
\end{equation}
where $P_{\ell}(\cos\beta)$ is the Legendre polynomial and $\cos\beta\equiv\bm{n}\cdot\bm{n'}.$
Note that after treating the anisotropy of the rotation angle as a Gaussian random field, the effects on CMB are totally presented in the rotated CMB power spectra. 
With these the rotated CMB power spectra have the following forms \cite{Li:2008tma, Li:2013vga}:
\begin{align}
&\widetilde{C}_\ell^{EE} + \widetilde{C}_\ell^{BB} = \exp[-4C^{\alpha}(0)]\sum_{\ell'}\frac{2\ell'+1}{2}(C_{\ell'}^{EE}+C_{\ell'}^{BB})\int_{-1}^1 d_{22}^{\ell'}(\beta) d_{22}^{\ell}(\beta) e^{4C^{\alpha}(\beta)} d\cos(\beta)\nn\\
&\widetilde{C}_\ell^{EE} - \widetilde{C}_\ell^{BB} = \cos(4\bar{\alpha})\exp[-4C^{\alpha}(0)]\sum_{\ell'}\frac{2\ell'+1}{2}(C_{\ell'}^{EE}-C_{\ell'}^{BB})\int_{-1}^1 d_{-22}^{\ell'}(\beta) d_{-22}^{\ell}(\beta) e^{-4C^{\alpha}(\beta)} d\cos(\beta)\nn\\
&\widetilde{C}_\ell^{EB}  = \sin(4\bar{\alpha})\exp[-4C^{\alpha}(0)]\sum_{\ell'}\frac{2\ell'+1}{4}(C_{\ell'}^{EE}-C_{\ell'}^{BB})\int_{-1}^1 d_{-22}^{\ell'}(\beta) d_{-22}^{\ell}(\beta) e^{-4C^{\alpha}(\beta)} d\cos(\beta)\nn\\
&\widetilde{C}_\ell^{TE}  = C_{\ell}^{TE}\cos(2\bar{\alpha})e^{-2C^{\alpha}(0)}\nn\\
&\widetilde{C}_\ell^{TB}  = C_{\ell}^{TE}\sin(2\bar{\alpha})e^{-2C^{\alpha}(0)}~,
\label{int_transform}
\end{align}
where $d_{22}^{\ell}$, $d_{-22}^{\ell}$ are the Wigner small d matrices.
Note that Eq. (\ref{int_transform}) reduces to the isotropic version as Eq.  (\ref{iso_rotation}) if the power spectrum of rotation angle $C_{\ell}^{\alpha\alpha}$ vanishes.

The integrals on the right hand side are computational expensive for large $\ell$. 
In order to speed up, we deduce the exponential terms in the Eq. (\ref{int_transform}) through Taylor series:
\begin{eqnarray}
e^{4C^{\alpha}(\beta) -4C^{\alpha}(0) } &=& 1 + [4 C^{\alpha}(\beta) - 4C^{\alpha}(0)]  + ...\nn \\
e^{-4C^{\alpha}(\beta) - 4C^{\alpha}(0) } &=& 1 - [4 C^{\alpha}(\beta)+ 4C^{\alpha}(0)] + ... ~.
\label{tylor_series}
\end{eqnarray}
Note that $\left|C^{\alpha}(\beta)\right|\le C^{\alpha}(0)$ since the absolute value of Legendre polynomial is always less than or equal to 1. 
According to the results presented in our previous work \cite{Li:2013vga, Li:2014oia}, the constraint on the variance $C^{\alpha}(0)<0.02$ at 95\% confidence level, so that both $C^{\alpha}(\beta)$ and $C^{\alpha}(0)$ can be treated as small variables.
Keep the expansion to the first order, the corresponding difference due to the polarization rotation can be expressed as:
\begin{eqnarray}
\widetilde{C}_{\ell}^{EE} &=& C_{\ell }^{EE} + \delta\widetilde{C}_{\ell, 0}^{EE}+ \delta\widetilde{C}_{\ell, 1 }^{EE} ,\nonumber\\ 
\widetilde{C}_{\ell}^{BB} &=& C_{\ell }^{BB} + \delta\widetilde{C}_{\ell, 0}^{BB} + \delta\widetilde{C}_{\ell, 1}^{BB} ,\\
\delta \widetilde{C}_{\ell, 0}^{EE} &=& -\delta\widetilde{C}_{\ell, 0}^{BB}  =  -(C_\ell^{EE} - C_\ell^{BB})\sin^2(2\bar{\alpha})~,
\label{split}
\end{eqnarray}
where $\delta\widetilde{C}_{l, 0}^{EE}, \delta\widetilde{C}_{l, 0}^{BB}$ represent the changes of the EE, BB power spectra due to the isotropic rotation angle $\bar{\alpha}$, and $\delta\widetilde{C}_{l, 1}^{EE}, \delta\widetilde{C}_{l, 1}^{BB}$ represent the variation induced by the anisotropic rotation $\delta{\alpha(\bm{n})}$.

Substituting Eq. (\ref{tylor_series}) into (\ref{int_transform}), we obtain the expression to the first order :
\begin{eqnarray}
\delta \widetilde{C}_{\ell, 1}^{EE} & =&  4C^\alpha(0)\bigg[ \sum_{\ell ''}  W^{\alpha}_{\ell ''}   \sum_{\ell'} M_{\ell, \ell ',\ell ''}  \begin{cases}
\widetilde{C}_{\ell', 0}^{EE},  & \ell_{\mbox{sum}} \mbox{ is odd}\\
\widetilde{C}_{\ell', 0}^{BB},  & \ell_{\mbox{sum}} \mbox{ is even}
\end{cases}- \widetilde{C}_{\ell, 0}^{EE} \bigg], \label{EE_1order}\\
\delta \widetilde{C}_{\ell, 1}^{BB} &=&   4C^\alpha(0)\bigg[ \sum_{\ell''}W^\alpha_{\ell''}  \sum_{\ell'} M_{\ell, \ell',\ell''} 
\begin{cases}
\widetilde{C}_{\ell', 0}^{BB},  &\ell_{\mbox{sum}} \mbox{ is odd}\\
\widetilde{C}_{\ell', 0}^{EE},  & \ell_{\mbox{sum}} \mbox{is even}.
\end{cases}-\widetilde{C}_{\ell, 0}^{BB} \bigg], \label{BB_1order}\\
\widetilde{C}_{\ell, 0}^{EE} &=&  C_\ell^{EE} + \delta \widetilde{C}_{\ell, 0}^{EE}, \\
\widetilde{C}_{\ell, 0}^{BB} &=&  C_\ell^{BB} + \delta \widetilde{C}_{\ell, 0}^{BB},
\end{eqnarray}
where $ \ell_{\mbox{sum}}= \ell + \ell' + \ell''$  ,  $W^{\alpha}_{\ell''}$ is weighting coefficient in terms of  power spectrum of the anisotropic rotation angle,  and $M_{\ell,\ell',\ell''}$ represents  weighting  factor composed by the square of  Wigner 3-j symbol, i.e., 
\begin{eqnarray}
W^\alpha_{\ell''} &=& \frac{2\ell''+1}{4\pi C^\alpha(0)} C_{\ell''}^{\alpha\alpha}, \quad \sum_{ \ell''} W^\alpha_{\ell''} = 1, \label{weight_alpha}\\
M_{\ell, \ell', \ell''} &=&(2\ell' +1) 
\left( {\begin{array}{ccc}
	\ell &\ell' & \ell'' \\
	2 & -2 & 0\\
	\end{array} } \right)^2, \quad \sum_{\ell'} M_{\ell, \ell', \ell''} = 1.\label{weight_3j}
\end{eqnarray}

\begin{figure} [htbp]
	\centering
	\subfigure[]{ 
		\label{fig_theory_eb}
		\includegraphics[width=3in]{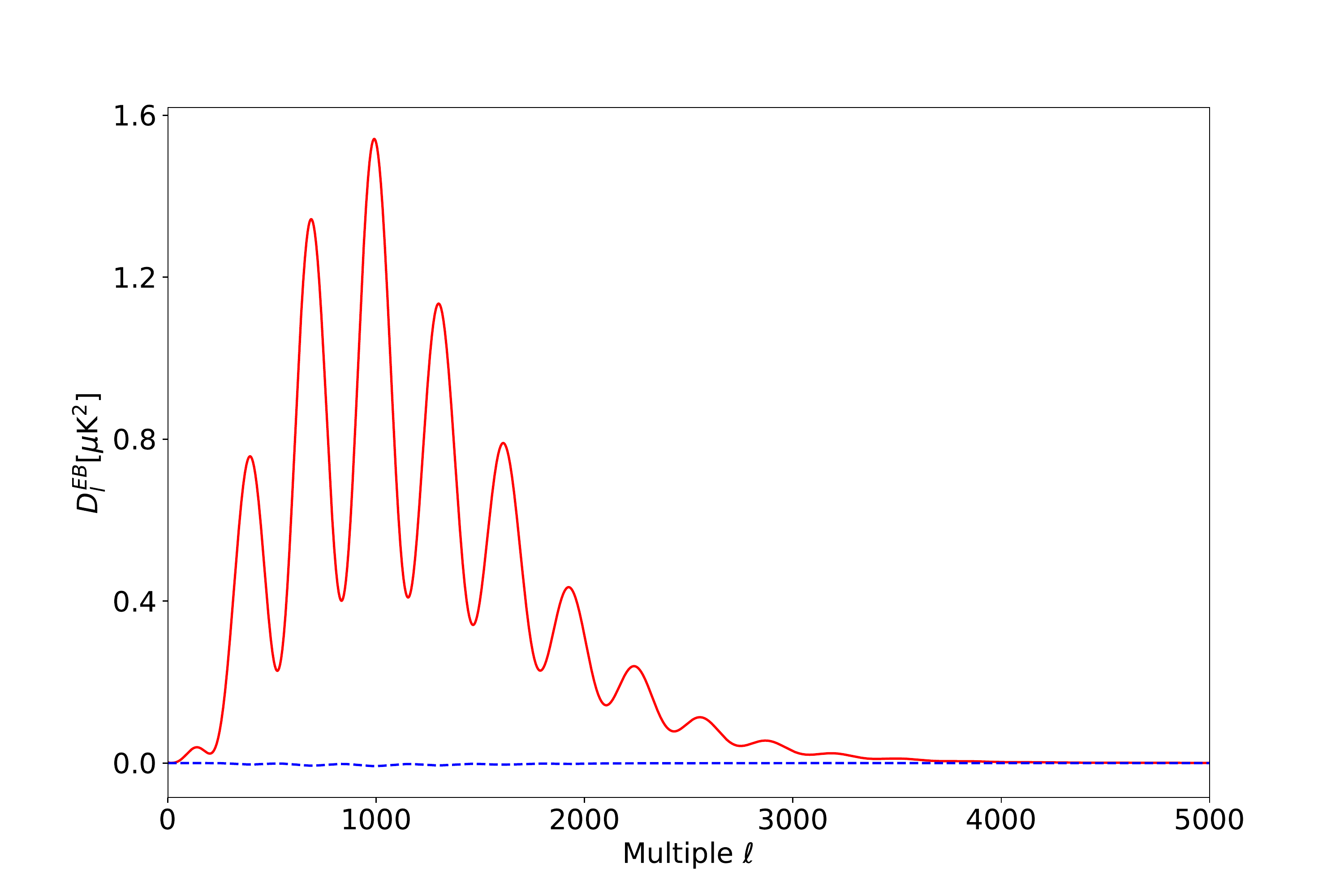} 
	}
    \subfigure[]{ 
	    \label{fig_theory_tb}
	    \includegraphics[width=3in]{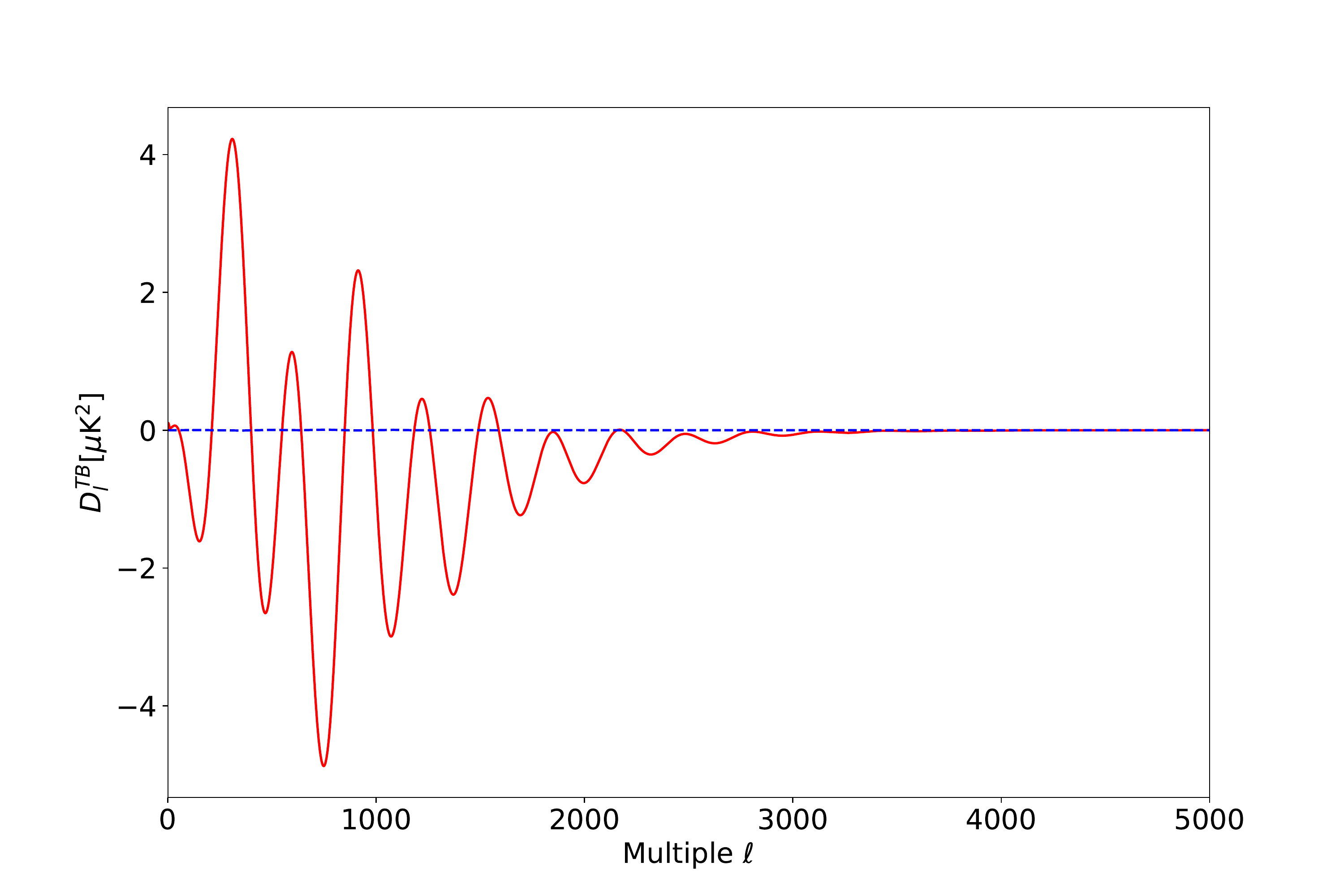} 
    }
	\subfigure[]{ 
		\label{fig_theory_ee}
		\includegraphics[width=3in]{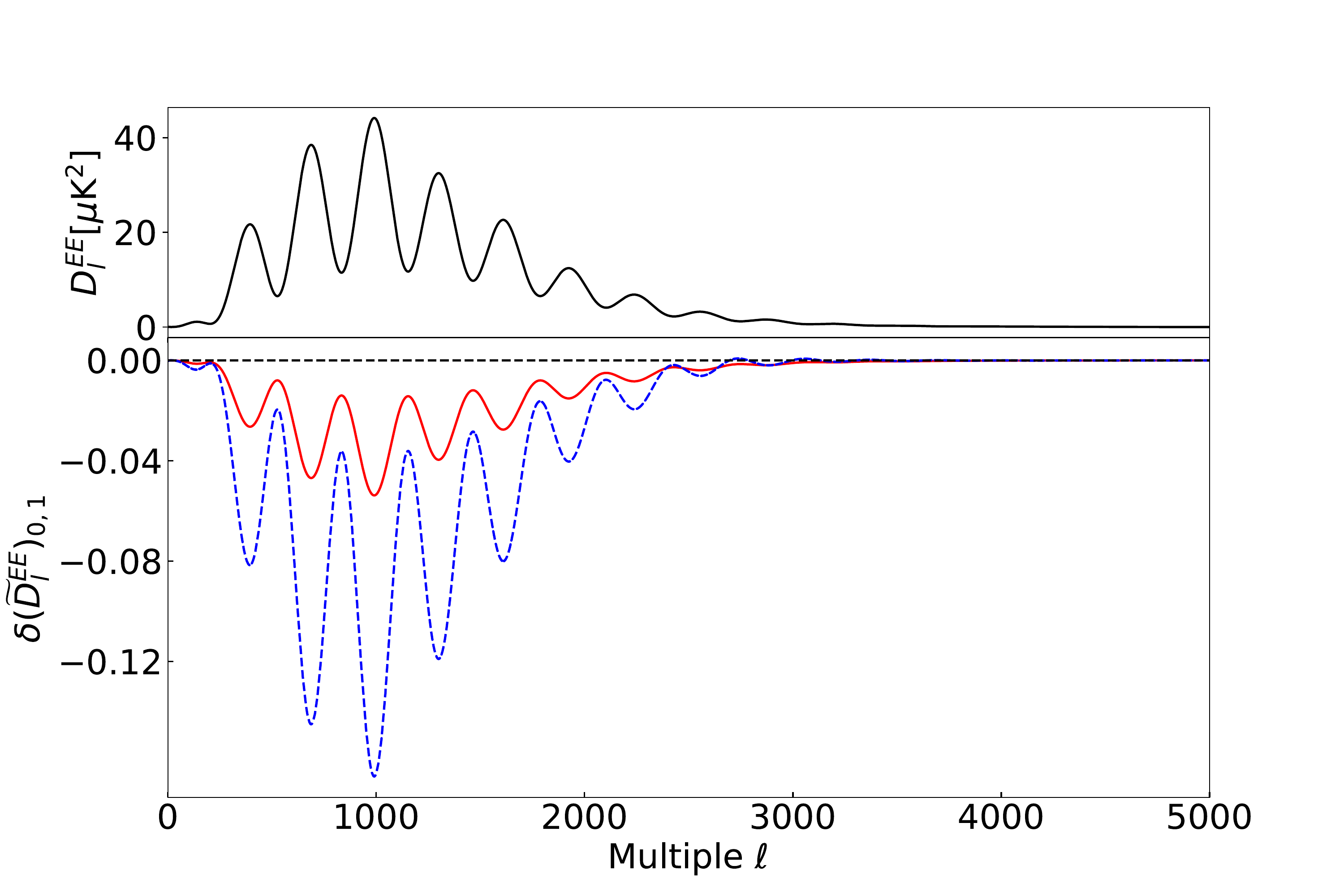} 
	} 
	\subfigure[]{ 
		\label{fig_theory_bb} 
		\includegraphics[width=3in]{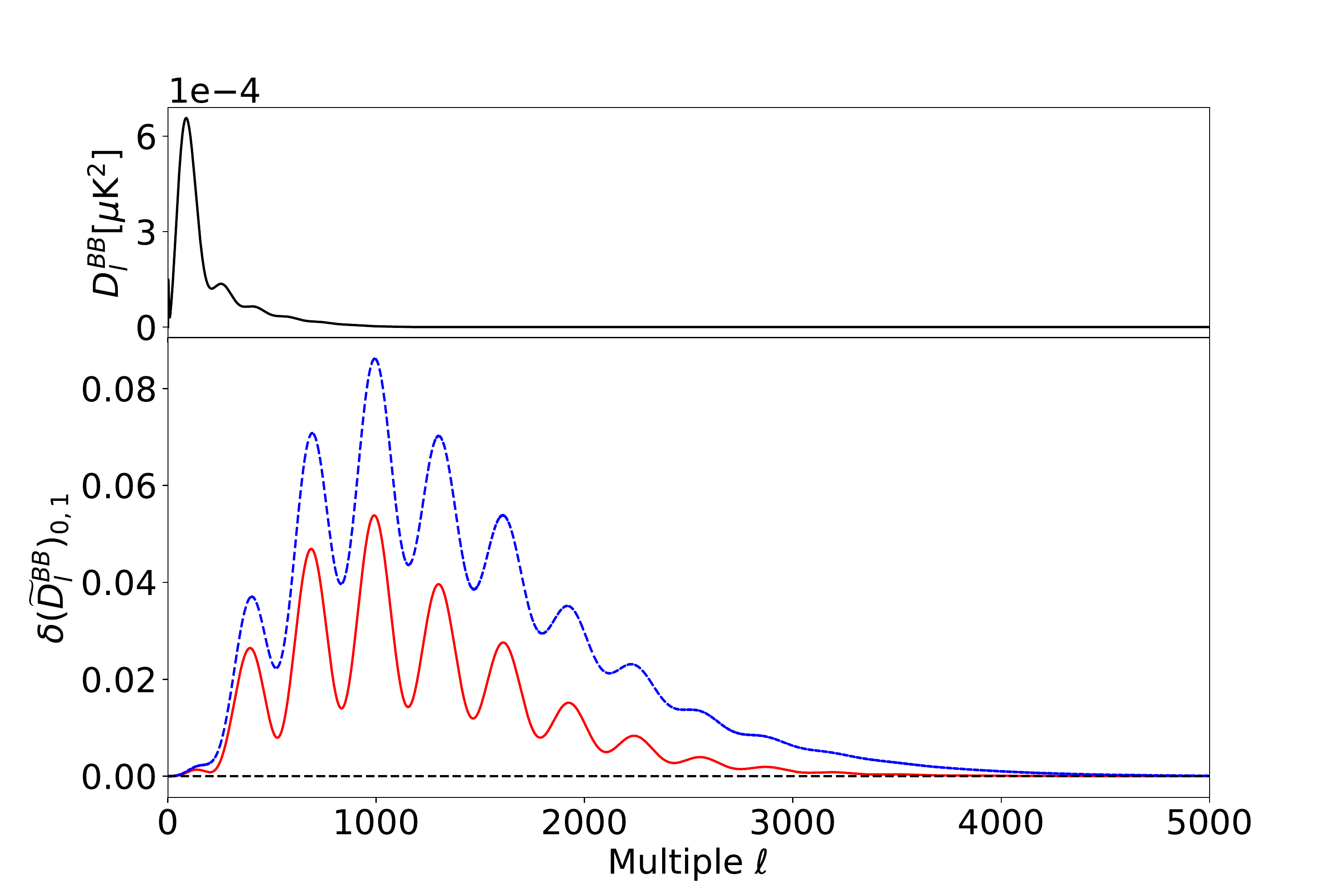} 
	} 
	\caption{ The effect of isotropic and anisotropic polarization rotation angle on  CMB  TB, EB, EE, BB power spectra. The solid black line represents the un-rotated spectra,  the solid red line corresponds to contribution from isotropic rotation angle, while dashed blue line  corresponds to contribution of anisotropic rotation angle.}   
	\label{theory_plot}
\end{figure}

Since the isotropic rotation angle is tightly constrained\cite{Aghanim:2016fhp}, generally we have $\widetilde{C}_{\ell, 0}^{BB} \ll  \widetilde{C}_{\ell, 0}^{EE}$. In that case,  $\widetilde{C}_{\ell, 0}^{EE}$ dominates the effect of anisotropic rotation angles in Eq. (\ref{BB_1order}) and $ \delta \widetilde{C}_{\ell, 1}^{BB}$  will be positive.
To see this intuitively, we plot the polarized CMB spectra in Fig. \ref{theory_plot} to illustrate the effect of the polarization rotation angle .
Here we consider a typical model with key parameters chosen as: the tensor-to-scalar ratio $r=0.01$, the isotropic rotation angle $\bar{\alpha} = 1^\circ$, and a scale invariant  power spectrum of anisotropic rotation angles $D_{\ell}^{\alpha\alpha}=\ell(\ell+1)/2\pi\cdot C_\ell^{\alpha\alpha}=1\times10^{-4}  \,\mbox{rad}^2 $ which corresponds to the  variance of rotation angle $C^{\alpha}(0) \simeq 0.001$ .
The solid black lines in both panels represent the un-rotated CMB power spectra, the solid red line and dashed blue lines represent $\delta\widetilde{C}_{\ell, 0}$ and $\delta\widetilde{C}_{\ell, 1}$, respectively. 
Generally, from the figures we can see that the isotropic part dominates the variation in EB/TB power spectra. 
The isotropic rotation angle reduce the EE spectrum and raise up the BB spectrum. The  anisotropic rotation angles also has this feature and furthermore it raises EE spectrum in the scale where EE spectrum approaches zero.
This is because the anisotropic rotation angles affects the CMB power spectra at larger multipole region than the isotropic rotation angle, which can be obviously seen by comparing Eqs. (\ref{split}), (\ref{EE_1order}), and (\ref{BB_1order}).
When it comes to the small scale, $\delta\widetilde{ C}_{\ell, 0}^{EE}$, $\delta \widetilde{C}_{\ell, 0}^{BB}$ damps to zero quickly since the isotropic rotation angle only coverts the E and B modes to each other at the same scale.
However, the contribution due to the anisotropic rotation angles is basically a convolution over all scales of the E, B modes.

\section{NUMERICAL ANALYSIS}
\label{section3}

We modify the public available Markov Chain Monte Carlo package for cosmological study CosmoMC \cite{Lewis:2002ah} and perform a global fitting analysis by combining current CMB measurements at all scales:
the Planck 2018 temperature and polarization data, specifically, containing high $\ell$ TT, TE, EE spectra with $30<\ell<2508$, low $\ell$ TT, TE, EE, BB spectra with $2<\ell<30$ and lensing data;
the BB spectrum of BK15 dataset with $30<\ell<600$;
the SPTpol dataset including the BB spectrum released in 2014 with $300<\ell<2300$, and the TE, EE spectra released in 2017 with $50<\ell<8000$;
the ACTPol dataset including the contamination-marginalized TT, TE, EE spectra with $350<\ell<4000$;
the POLARBEAR BB spectrum released in 2014 with $500<\ell<2100$.  We neglect any correlation between these datasets.

In our fittings, the basic parameter space is defined as $P_{\Lambda\mathrm{CDM}} + r\equiv(\omega_b, \omega_c, \Omega_{\Lambda}, \tau , A_s, n_s, A_{\mathrm{lens}}, r)$.
In order to investigate the influence induced by the polarization rotation, we further consider two types of extension to the basic parameter space as $P_{\Lambda\mathrm{CDM}} + r + \bar{\alpha}$ and $P_{\Lambda\mathrm{CDM}} + r + \bar{\alpha} + C^{\alpha}(0) + D_{\ell}^{\alpha\alpha}$, where the $\omega_b\equiv\Omega_b h^2$ is the physical baryon density relative to the critical density, the $\omega_c\equiv\Omega_c h^2$, $\Omega_{\Lambda}$ are the ones to the cold dark matter and dark energy, $\tau$ is the optical depth during the reionization epoch, $A_s$ and $n_s$ are the amplitude and spectral index of the primordial scalar perturbation, $r$ is the tensor-to-scalar ratio which characterizes the amplitude of the primordial gravitational waves, $A_{\mathrm{lens}}$ is the amplitude of the lensing spectrum, and we set the pivot scale of primordial spectra as $k=0.05~\mathrm{Mpc}^{-1}$. 
For the additional parameters of CPT-violating effect, $\bar{\alpha}$ means the isotropic rotation angle, $D_{\ell}^{\alpha\alpha}$ is the power spectrum of anisotropic rotation angles defined as $\ell(\ell+1)/2\pi\cdot C_\ell^{\alpha\alpha}$. 

We take six parameters $ D_{\ell}^{\alpha\alpha}(i), i\in[1, 6]$, as the binned power spectrum, where each parameter represents the average value within the multipole range of [2, 350), [350, 700), [700, 1100), [1100, 1500), [1500, 2000), [2000, 2500]. Note that we neglect power spectrum of anisotropic rotation angles with multipole $\ell $ beyond 2500.
$C^{\alpha}(0)$ is the variance of anisotropic fluctuation of rotation angles, and we treat it as an independent parameter here following the ways in the papers \cite{Li:2013vga, Li:2014oia}. Naturally, $C^{\alpha}(0),  D_{\ell}^{\alpha\alpha}(i)$ are set to be positive definite parameters in this work.

\begin{figure}[htbp]
	\centering
	\includegraphics[width=5in]{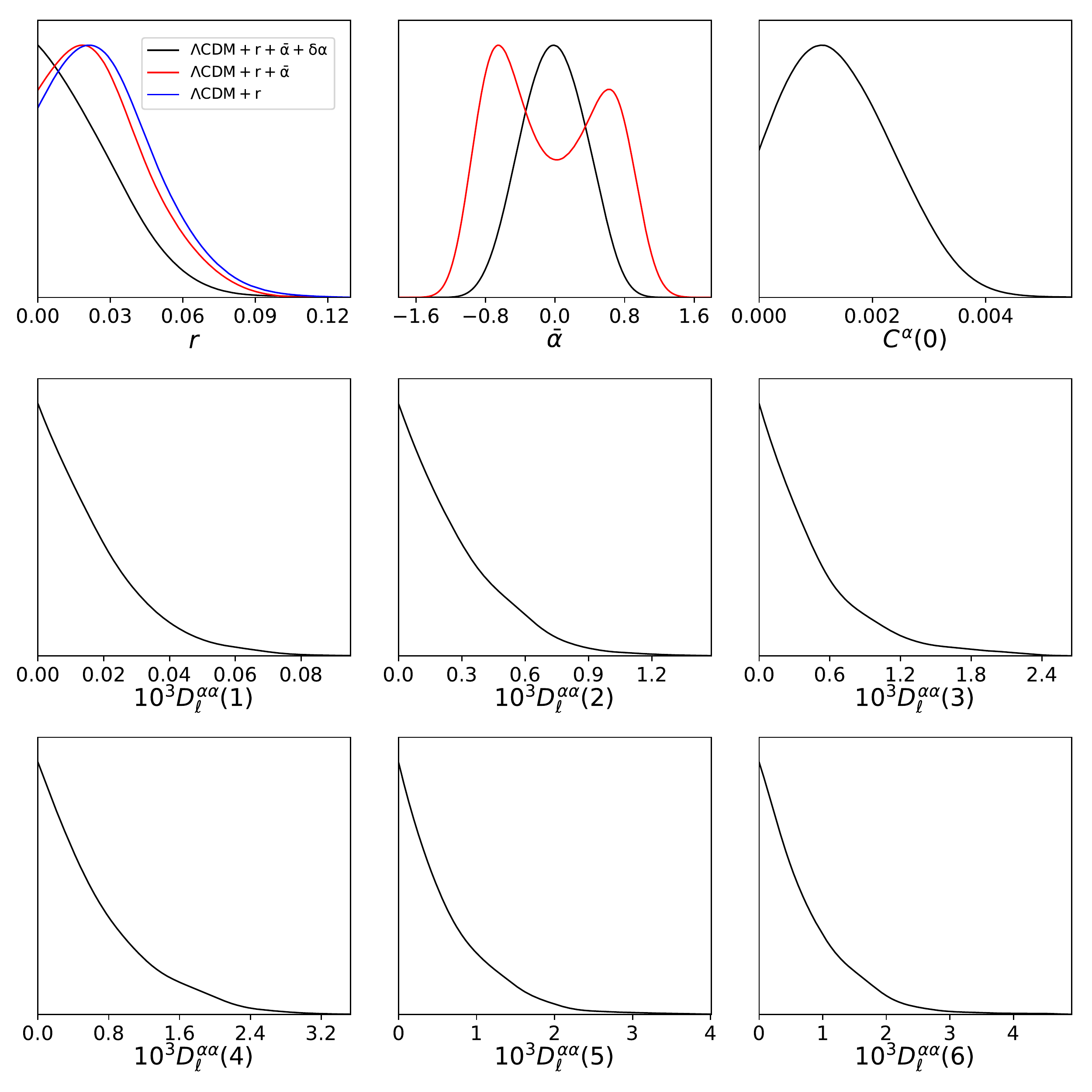}
	\caption{One dimensional posterior distributions of tensor-to-scalar ratio $r$ as well as isotropic/anisotropic rotation angles. The blue line is the constraint for $\mathrm{\Lambda CDM} $ model with $r$ extension,   the red line is for gravitational wave and isotropic rotation angle extension, and black line represents for both gravitational wave and isotropic/anistropic rotation angle extension. The unit of X-axis for the plot of $\bar{\alpha}$ is degree, and the ones for the plots of $D_{\ell}^{\alpha\alpha}$ are radians square.}
	
	\label{fitting_figure}	
\end{figure}

We list the constraints on tensor-to-scalar ratio and polarization rotation parameters in Table \ref{fitting_table}.
Relevant one dimensional distributions are also plotted in Fig. \ref{fitting_figure}.
For the other parameters of the canonical $\Lambda\mathrm{CDM}$ model, we do not find any significant change on their constraints whether the polarization rotation effect is considered or not.

\begin{table}[ht]
	\caption{$2\sigma$ Constraints on $r$ and polarization rotation angles.}
	\centering
	\begin{tabular}{p{3cm}<{\centering}|p{3cm}<{\centering}|p{3cm}<{\centering}|p{3cm}<{\centering}} 
		\hline  
		\hline 
		- & $P_{\Lambda\mathrm{CDM}} + r$& $P_{\Lambda\mathrm{CDM}} + r + \bar{\alpha}$  &  $P_{\Lambda\mathrm{CDM}} + r + \bar{\alpha} + D_{\ell}^{\alpha\alpha}(\, \mathrm{rad}^2)$ \\
		\hline  
		$r$ & $<0.069$  & $<0.067$ &    $ <0.057$ \\
		$\bar{\alpha}$ & - & $-0.1^\circ\pm 1.0^\circ$ &  $-0.01^\circ\pm 0.70^\circ$ \\
		$D_{\ell}^{\alpha\alpha}(1) $ & - & - &  $<4.71\times 10^{-5}$ \\
		$D_{\ell}^{\alpha\alpha}(2)$& -& - &  $<7.13\times 10^{-4}$ \\
		$D_{\ell}^{\alpha\alpha}(3)$& -& - &  $<1.35\times 10^{-3}$ \\
		$D_{\ell}^{\alpha\alpha}(4)$& -& - &  $<1.85\times 10^{-3}$ \\
		$D_{\ell}^{\alpha\alpha}(5)$& -& - &  $<1.83\times 10^{-3}$ \\
		$D_{\ell}^{\alpha\alpha}(6)$& - & - & $<2.08\times 10^{-3}$ \\
		$C^{\alpha}(0)$& - & - &  $<0.0032$\\
		\hline 
	\end{tabular}
	\label{fitting_table}
\end{table}

In order to validate our code, we apply the $P_{\Lambda\mathrm{CDM}} + r$ model in the first place, and obtain the constraint on r: $r<0.069$ at 95\% confidence level, which is  consistent with that derived from the BICEP2/Keck collaboration\cite{Ade:2018gkx}.

Similarly, we perform the global fitting using the same CMB datasets but considering polarization rotation effects induced by  Chern-Simons interaction.
We first concentrate on the derived limits of polarization rotation parameters.  
For the isotropic rotation angle, we find the CMB datasets can hardly constrain it very well, since the isotropic angle mainly causes deformation of CMB TB/EB spectra and the datasets we considered lacks in these regards.
We obtain the $1\sigma$ constraint on $\bar{\alpha}=-0.06^\circ\pm0.64^\circ$ under the $P_{\Lambda\mathrm{CDM}} + r + \bar{\alpha}$ model, and $\bar{\alpha}=-0.01\pm0.37$ when the anisotropic rotation is also considered.
Both of them agree with the result of Planck\cite{Aghanim:2016fhp}. 

For the constraints on the anisotropic rotation angles, 
we obtain the constraint on the variance of rotation angle $C^{\alpha}(0)<0.003$ at 95\% confidence level, which improves the result in Ref. \cite{Li:2013vga, Li:2014oia} by one order of magnitude.
Meanwhile, the upper limitations on binned power spectra of anisotropic rotation angles $C^{\alpha\alpha}_{\ell}$ we derived here will reach $10^{-9} \sim 10^{-8}$ after taking the factor $\ell(\ell+1)/2\pi$ off, which enhanced the results reported in \cite{Li:2014oia} by one order of magnitude.
These improvements are primarily from the CMB datasets with high $\ell$, such as POLARBEAR, ACTPol and SPTPol.
Anisotropic polarization rotation tends to increase the CMB BB spectrum over almost all multipoles, so that when it comes to the CMB data at small scale, the anisotropic angles will be constrained much tighter than the isotropic one.

So far as the constraint on r is concerned, we find that the $2\sigma$ upper limit of $r$ shrinks to $r<0.057$ when including the anisotropic polarization rotation effect.
More interestingly, from the probability distribution functions of $r$ in Fig. \ref{fitting_figure},  
we can see that the peak shifts to left and finally disappears due to the polarization rotation, since the polarization rotation angle shares a cut of B mode which is thought to be originated by primordial tensor perturbation.
This phenomena attributes to  our previous research\cite{Li:2015vea} on the degeneracy between r and polarization rotation angle, that it is necessary to consider the potential impact induced by the rotation angle if we attempts to pick up the weak signal of primordial tensor perturbation from CMB BB spectrum.

\section{Summary}

In this work we performed a global fitting on CMB B mode parameters\cite{xmzhang:closing} with latest CMB  data from Planck, BICEP2/Keck Array,  POLARBEAR, SPTpol, and ACTPol.  We obtained the isotropic rotation angle $\bar{\alpha} = -0.01^\circ \pm 0.37^\circ $ at 68\% C.L,
and the 95\% C.L. upper limit on the variance of the anisotropic rotation angles $C^{\alpha}(0)<0.0032~\mathrm{rad}^2$.  We binned the power spectrum of polarization rotation angle and obtained the constraint  $D^{\alpha\alpha}_{\ell\in [2,350]}<4.71\times10^{-5}\rm{rad}^2$.
Meanwhile, after considering the rotation angle effects, we found that the  one dimensional posterior distribution of $r$ is changed , and its 95\% C.L. upper limit gets tightened from $0.069$ to $0.057$.

\label{summary}

\begin{acknowledgments}
We thank Yang Liu, Chang Feng and Gongbo Zhao for useful discussions.  H. Z., S. L. and X. Z. are supported in part by the Ministry of Science and Technology of China (2016YFE0104700), the National Natural Science Foundation of China (Grant NO. 11653001), the CAS pilot B project (XDB23020000).  M. L. is supported by NSFC under Grant No. 11653002. 
\end{acknowledgments}


\begin{thebibliography}{00}
	
\bibitem{Li:2001st} 
Mingzhe Li, Xiulian Wang, Bo Feng and Xinmin Zhang,
Phys.\ Rev.\ D {\bf 65}, 103511 (2002)
[hep-ph/0112069].

\bibitem{Li:2002wd} 
Mingzhe Li and Xinmin Zhang,
Phys.\ Lett.\ B {\bf 573}, 20 (2003)
[hep-ph/0209093].

\bibitem{Davoudiasl:2004gf} 
Hooman Davoudiasl, Ryuichiro Kitano, Graham D. Kribs, Hitoshi Murayama and Paul J. Steinhardt,
Phys.\ Rev.\ Lett.\  {\bf 93}, 201301 (2004)
[hep-ph/0403019].

\bibitem{Li:2004hh} 
Hong Li, Mingzhe Li, and Xinmin Zhang,
Phys.\ Rev.\ D {\bf 70}, 047302 (2004)
[hep-ph/0403281].

\bibitem{Cohen:1987vi} 
Andrew G. Cohen and  David B. Kaplan 
Phys.\ Lett.\ B {\bf 199}, 251 (1987).

\bibitem{Feng:2004mq} 
Bo Feng, Hong Li, Mingzhe Li, and Xinmin Zhang,
Phys.\ Lett.\ B {\bf 620}, 27 (2005)
[hep-ph/0406269].

\bibitem{Feng:2006dp} 
Bo Feng, Mingzhe Li, Jun-Qing Xia, Xuelei Chen, and Xinmin Zhang,
Phys.\ Rev.\ Lett.\  {\bf 96}, 221302 (2006)
[astro-ph/0601095].

\bibitem{Li:2006ss} 
Mingzhe Li, Jun-Qing Xia, Hong Li, and Xinmin Zhang,
Phys.\ Lett.\ B {\bf 651}, 357 (2007)
[hep-ph/0611192].

\bibitem{Li:2009rt} 
Mingzhe Li, Yi-Fu Cai, Xiulian Wang, and Xinmin Zhang,
Phys.\ Lett.\ B {\bf 680}, 118 (2009)
doi:10.1016/j.physletb.2009.08.053
[arXiv:0907.5159 [hep-ph]].

\bibitem{Wetterich:1987fm} 
C.~Wetterich,
Nucl.\ Phys.\ B {\bf 302}, 668 (1988)
[arXiv:1711.03844 [hep-th]].

\bibitem{Caldwell:1999ew} 
R.~R.~Caldwell,
Phys.\ Lett.\ B {\bf 545}, 23 (2002)
[astro-ph/9908168].

\bibitem{Feng:2004ad} 
Bo Feng, Xiulian Wang, and Xinmin Zhang,
Phys.\ Lett.\ B {\bf 607}, 35 (2005)
[astro-ph/0404224].


\bibitem{Li:2008tma} 
Mingzhe Li and Xinmin Zhang,
Phys.\ Rev.\ D {\bf 78}, 103516 (2008)
[arXiv:0810.0403 [astro-ph]].


\bibitem{Lue:1998mq} 
Arthur Lue, Limin Wang and Marc Kamionkowski,
Phys.\ Rev.\ Lett.\  {\bf 83}, 1506 (1999)
[astro-ph/9812088].




\bibitem{Wu:2008qb} 
E.~Y.~S.~Wu {\it et al.} [QUaD Collaboration],
Phys.\ Rev.\ Lett.\  {\bf 102}, 161302 (2009)
[arXiv:0811.0618 [astro-ph]].

\bibitem{Hinshaw:2012aka} 
G.~Hinshaw {\it et al.} [WMAP Collaboration],
Astrophys.\ J.\ Suppl.\  {\bf 208}, 19 (2013)
[arXiv:1212.5226 [astro-ph.CO]].

\bibitem{Louis:2016ahn} 
Thibaut Louis{\it et al.} [ACTPol Collaboration],
JCAP {\bf 1706}, 031 (2017)
[arXiv:1610.02360 [astro-ph.CO]].

\bibitem{Aghanim:2016fhp} 
N.~Aghanim {\it et al.} [Planck Collaboration],
Astron.\ Astrophys.\  {\bf 596}, A110 (2016)
[arXiv:1605.08633 [astro-ph.CO]].

\bibitem{Xia:2009ah} 
Jun-Qing Xia, Hong Li, and Xinmin Zhang,
Phys.\ Lett.\ B {\bf 687}, 129 (2010)
[arXiv:0908.1876 [astro-ph.CO]].

\bibitem{Zhao:2015mqa} 
Gong-Bo Zhao, Yuting Wang, Jun-Qing Xia, Mingzhe Li, and Xinmin Zhang,
JCAP {\bf 1507}, 032 (2015)
[arXiv:1504.04507 [astro-ph.CO]].

\bibitem{Xia:2012ck} 
Jun-Qing Xia,
JCAP {\bf 1201}, 046 (2012)
[arXiv:1201.4457 [astro-ph.CO]].

\bibitem{Li:2013vga} 
Mingzhe Li and Bo Yu,
JCAP {\bf 1306}, 016 (2013)
[arXiv:1303.1881 [astro-ph.CO]].

\bibitem{Li:2014oia} 
Si-Yu Li, Jun-Qing Xia, Mingzhe Li, Hong Li, and Xinmin Zhang,
Astrophys.\ J.\  {\bf 799}, no. 2, 211 (2015)
[arXiv:1405.5637 [astro-ph.CO]].


\bibitem{Kamionkowski:2008fp} 
Marc Kamionkowski, 
Phys.\ Rev.\ Lett.\  {\bf 102}, 111302 (2009)
[arXiv:0810.1286 [astro-ph]].

\bibitem{Yadav:2009eb} 
Amit P.S. Yadav, Rahul Biswas, Meng Su, and Matias Zaldarriaga,
Phys.\ Rev.\ D {\bf 79}, 123009 (2009)
[arXiv:0902.4466 [astro-ph.CO]].

\bibitem{Gluscevic:2009mm} 
Vera Gluscevic, Marc Kamionkowski, and Asantha Cooray,
Phys.\ Rev.\ D {\bf 80}, 023510 (2009)
[arXiv:0905.1687 [astro-ph.CO]].


\bibitem{Gluscevic:2012me} 
Vera Gluscevic, Duncan Hanson, Marc Kamionkowski and Christopher M. Hirata,
Phys.\ Rev.\ D {\bf 86}, 103529 (2012)
[arXiv:1206.5546 [astro-ph.CO]].



\bibitem{Ade:2015cao} 
P.~A.~R.~Ade {\it et al.} [POLARBEAR Collaboration],
Phys.\ Rev.\ D {\bf 92}, 123509 (2015)
[arXiv:1509.02461 [astro-ph.CO]].


\bibitem{Array:2017rlf} 
P.~A.~R.~Ade {\it et al.} [BICEP2 and Keck Arrary Collaborations],
Phys.\ Rev.\ D {\bf 96}, no. 10, 102003 (2017)
[arXiv:1705.02523 [astro-ph.CO]].

\bibitem{Li:2014cka} 
Hong Li, Jun-Qing Xia and Xinmin Zhang,
arXiv:1404.0238 [astro-ph.CO].

\bibitem{Li:2015vea} 
Si-Yu Li, Jun-Qing Xia, Mingzhe Li and Xinmin Zhang,
Phys.\ Lett.\ B {\bf 751}, 579 (2015)
[arXiv:1506.03526 [astro-ph.CO]].

\bibitem{Ade:2018gkx} 
P.~A.~R.~Ade {\it et al.} [BICEP2 and Keck Array Collaborations],
Phys.\ Rev.\ Lett.\  {\bf 121}, 221301 (2018)
[arXiv:1810.05216 [astro-ph.CO]].

\bibitem{Aghanim:2018eyx} 
N.~Aghanim {\it et al.} [Planck Collaboration],
arXiv:1807.06209 [astro-ph.CO].

\bibitem{Ade:2014afa} 
P.~A.~R.~Ade {\it et al.} [POLARBEAR Collaboration],
Astrophys.\ J.\  {\bf 794}, no. 2, 171 (2014)
doi:10.1088/0004-637X/794/2/171
[arXiv:1403.2369 [astro-ph.CO]].


\bibitem{Henning:2017nuy} 
J.~W.~Henning {\it et al.} [SPT Collaboration],
Astrophys.\ J.\  {\bf 852}, no. 2, 97 (2018)
[arXiv:1707.09353 [astro-ph.CO]].

\bibitem{Keisler:2015hfa} 
R.~Keisler {\it et al.} [SPT Collaboration],
Astrophys.\ J.\  {\bf 807}, no. 2, 151 (2015)
[arXiv:1503.02315 [astro-ph.CO]].




\bibitem{Lewis:2002ah} 
Antony Lewis and Sarah Bridle,
Phys.\ Rev.\ D {\bf 66}, 103511 (2002)
[astro-ph/0205436].

\bibitem{xmzhang:closing}
Xinmin Zhang, closing remark at the The 2nd International Symposium on Cosmology and Ali CMB Polarization Telescope, September 7-9, 2019 Beijing, China. 
\end{thebibliography}
\end{document}